\documentclass[10pt, conference, letterpaper]{IEEEtran}

\usepackage{booktabs} 
\usepackage{pifont}		
\usepackage{graphicx}
\usepackage{epstopdf}
\usepackage{enumitem}
\usepackage{epstopdf}
\usepackage{graphicx}
\usepackage{amsmath}
\usepackage{subcaption}
\usepackage{comment}
\usepackage[english]{babel}
\usepackage[utf8]{inputenc}
\usepackage{algorithm}
\usepackage{algpseudocode}
\usepackage{amsfonts}
\usepackage{amssymb}
\usepackage{setspace}
\usepackage{subcaption}
\usepackage{cite}

\ifCLASSINFOpdf
\else
\fi

\begin{document}

\title{Energy-Efficient Mobile Network I/O}

\author{\IEEEauthorblockN{Kemal Guner and Tevfik Kosar}
	\IEEEauthorblockA{Department of Computer Science and Engineering\\
		University at Buffalo, (SUNY), Buffalo, NY 14260, USA\\
		Email: \{kemalgne, tkosar\}@buffalo.edu}
}

\maketitle

\thispagestyle{empty}
\pagestyle{plain}

\begin{abstract}
By year 2020, the number of smartphone users globally will reach 3 Billion and the mobile data traffic (cellular + WiFi) will exceed PC Internet traffic the first time. 
As the number of smartphone users and the amount of data transferred per smartphone grow exponentially, limited battery power is becoming an increasingly critical problem for mobile devices which heavily depend on network I/O. Despite the growing body of research in power management techniques for the mobile devices at the hardware layer as well as the lower layers of the networking stack, there has been little work focusing on saving energy at the application layer for the mobile systems during network I/O. 
In this paper, we show that 
significant energy savings can be achieved with application-layer solutions at the mobile systems during data transfer with no performance penalty. In many cases, performance increase and energy savings can be achieved simultaneously.
\end{abstract}

\IEEEpeerreviewmaketitle
\vspace{3mm}
\renewcommand\IEEEkeywordsname{Keywords }

\section{Introduction}
\label{sec:intro}

The number of smartphone users globally has already exceeded 2 Billion, and this number is expected to reach 3 Billion by 2020~\cite{edwards2016gamification}. It is also estimated that smartphone mobile data traffic (cellular + WiFi) will reach 370 Exabytes per year by that time, exceeding PC Internet traffic the first time in the history~\cite{Cisco_2016}. 
An average smartphone consumes between 300 -- 1200 milliwatts power~\cite{carroll2010analysis} depending on the type of applications it is running, and most of the energy in smartphone applications is spent for network I/O. During an active data transfer, the cellular and WiFi components of a smartphone consume more power than its CPU, RAM, and even LCD+graphics card at the highest brightness level \cite{pathak2012energy, carroll2010analysis}.  Although the mobile data traffic and the amount of energy spent for it increase at a very fast pace, the battery capacities of smartphones do not increase at the same rate.

Limited  battery  power  is becoming an increasingly critical  problem  for smartphones and mobile computing, and many techniques have been proposed in the literature to overcome this at different layers. 
At the physical layer, techniques were proposed to choose appropriate modulation, coding, and transmission power control schemes to improve energy efficiency of the mobile device~\cite{cianca2001improving, schurgers2001modulation}.
At the media access control (MAC) layer, several new energy-efficient MAC protocol designs were proposed~\cite{ye2002energy, bharghavan1994macaw}.
At the network layer, low-power and scalable routing algorithms were developed~\cite{ chang2000energy, seada2004energy}.
At the transport layer, traffic shaping techniques and new transport protocols~\cite{kravets2000application, akella2001protocols} were proposed to exploit application-specific information and reduce power utilization.  

Despite the growing body of research in power management techniques for the lower layers of the mobile networking stack, there has been little work focusing on saving network I/O (data transfer) energy at the application layer. 
The most notable work in this area are: tuning the client playback buffer size during media streaming in order to minimize the total energy spent~\cite{bertozzi2002power}; using lossless compression techniques to minimize the amount of data transferred as well as the energy consumed on wireless devices ~\cite{xu2003impact}; and joint optimization of the application layer, data link layer, and physical layer of the protocol stack using an application-oriented objective function in order to improve multimedia quality and power consumption at the same time~\cite{khan2006application}.

In this paper, we show that significant amount of network I/O energy savings can be obtained at the application layer with no performance penalty. 
Application-layer optimization has the benefit of not requiring any changes to the smartphone hardware, to the operating system kernel, or to the lower-layer networking stack, although its deployment at scale will be very easy and its impact will be very big considering the end-to-end performance of the mobile network I/O and its energy efficiency will increase drastically.
We analyze the effects of different application layer data transfer protocol parameters (such as the number of parallel data streams per file, and the level of concurrent file transfers) on mobile data transfer throughput and energy consumption. 

In summary, the contributions of this paper are as follows: 

\begin{itemize}[leftmargin=*]
	\vspace{-1mm}
	\item To the best of our knowledge, we are first to provide an in depth analysis of the effects of application layer data transfer protocol parameters on the energy consumption of mobile phones.
	We show that significant energy savings can be achieved with application-layer solutions at the mobile systems during data transfer with no performance penalty.
	We also show that, in many cases, performance increase and energy savings can be achieved simultaneously.
	
	\item Worldwide energy consumption by smartphones is expected to be 24 terawatt hours in 2020~\cite{carroll2010analysis, Cisco_2016}, which translates to around 5 billion U.S. dollars per year, with a U.S. share of 10\%~\cite{statista2016}. This work will help to reduce the total energy consumption cost of smartphones worldwide through a fully application-layer solution which will be very easy to deploy at a large-scale. Our preliminary work shows that we can save up to 81\% energy using application-layer techniques in certain cases. 

	\item The majority of the mobile users fail to obtain even a fraction of the theoretical speeds promised by the existing mobile networks due to sub-optimal transport protocol tuning. 
	This work will help increasing the mobile data transfer speed by efficiently tuning application-layer data transport parameters. Our preliminary work shows that we can achieve up to 8.5X performance improvement while saving energy in certain cases. 
	
\end{itemize}

The rest of this paper is organized as follows: Section II provides background information on energy-aware tuning of application-layer data transfer protocol parameters and discusses the related work in this area; Section III explains the details of experimental setup and the analysis of individual network tuning parameters; Section IV introduces our three novel algorithms for mobile data transfer; Section V presents the results of our algorithms and comparison of them with other standard applications on mobile data transfer performance and energy consumption; and Section VI concludes the paper.

\section{Energy-Efficient Mobile Networking}
\label{sec:background}

The majority of work on mobile device energy savings focuses putting the devices to sleep during idle times~\cite{vallina2011erdos, vallina2013energy}. A recent study by Dogar et al.~\cite{Dogar2010} takes this approach to another step, and puts the device into sleep even during data transfer by exploiting the high-bandwidth wireless interface. They combine small gaps between packets into meaningful sleep intervals, thereby allowing the NIC as well as the device to doze off. Another track of study in this area focuses on switching among multiple radio interfaces in an attempt to reduce the overall power consumption of the mobile device~\cite{balasubramanian2009energy, nika2015energy}.
These techniques are orthogonal to our application-layer protocol tuning approach and could be used together to achieve higher energy efficiency in the mobile systems.

The closest work to ours in the literature is the work by Bertozzi et al.~\cite{bertozzi2003transport}, in which they investigate the energy trade-off in mobile networking as a function of the TCP receive buffer size and show that the TCP buffering mechanisms can be exploited to significantly increase energy efficiency of the transport layer with minimum performance overheads.

In this work, we focus on the tuning of two important protocol parameters:  concurrency (the level of concurrent file transfers to fill the mobile network pipes),
and parallelism (the number of parallel data streams per file).
{\em Concurrency} refers to sending multiple files simultaneously through the network using different data channels at the same time~\cite{kosar04, Thesis_2005, Kosar09, JGrid_2012}.
{\em Parallelism} sends different chunks of the same file using different data channels (i.e., TCP streams) at the same time 
and achieves high throughput by mimicking the behavior of individual streams and getting a higher share of the available bandwidth~\cite{R_Sivakumar00, R_Lee01, R_Balak98, R_Hacker05, R_Eggert00, R_Karrer06, R_Lu05, DADC_2008, DADC_2009, NDM_2011}.
Predicting the optimal concurrency and parallelism numbers for a specific setting is a very challenging problem due to the dynamic nature of the interfering background traffic~\cite{TCP_Pipeline, farkas2002, Grid_2008}. Using too many simultaneous connections would congest the network and the throughput will start dropping down.

When used wisely, these parameters have a potential to improve the end-to-end data transfer performance at a great extent, but improper use of these parameters can also hurt the performance of the data transfers due to increased load at the end-systems and congested links in the network. For this reason, it is crucial to find the best combination for these parameters with the least intrusion and overhead to the system resource utilization and power consumption.

In the literature, several highly-accurate predictive models ~\cite{R_Yin11, R_Yildirim11, DISCS12, Cluster_2015} were developed which would require as few as three sampling points to provide very accurate predictions for the parallel stream number giving the highest transfer throughput for the wired networks. 
Yildirim et al. analyzed the combined effect of parallelism and concurrency on end-to-end data transfer throughput~\cite{TCC_2016}. 
Managed File Transfer (MFT) systems were proposed which used a subset of these parameters in an effort to improve the end-to-end data transfer throughput~\cite{WORLDS_2004, ScienceCloud_2013, globusonline, Royal_2011, IGI_2012}.
Engin et al.\cite{EnginHARP2016} and Nine et al.\cite{NineBigData2017} proposed state-of-the-art algorithms that take into account both historical data analysis and dynamic tuning of application-layer parameters.
Alan et al.~\cite{Alan2015,Kosar_jrnl14} analyzed the effects of parallelism and concurrency on end-to-end data transfer throughput versus total energy consumption in wide-area wired networks in the context of GridFTP data transfers~\cite{R_Allcock05, NDM_2012}. {\em None of the existing work in this area studied the effects of these parameters on the mobile energy consumption and the performance versus energy trade-offs of tuning these parameters in this context.}

\section{Analysis of Parameter Effects}
\label{sec:experiments}

In our analysis, we have used a single-phase portable Yokogawa WT210 power meter, which provides highly accurate and fine granular power values (up to 10 readings per second) and is one of the accepted devices by the Standard Performance Evaluation Corporation (SPEC) power committee for power measurement and analysis purposes in the field. This power meter is used to measure the power consumption rates during the data transfers at the mobile client device. 

We designed a testbed with four different mobile devices, which are Google Nexus S, Samsung Galaxy Nexus N3, Galaxy S4, and Galaxy S5. We tested both WiFi and 4G LTE connections in progress of data transfers on end-systems. To reduce the effect of number of active users and the effect of peak/off-peak hours during the transfers, we adopted a strategy of using different time windows for each run of the same experiment setting, and took the average throughput and energy consumption values. We conducted all experiments at the same location and with the same distance and interference for objective analysis of the end-system devices. 

We choose HTTP (Hypertext Transport Protocol) as the application-layer transfer protocol to test the impact of the parameters of interest on the end-to-end data transfer throughput as well as the energy consumption of the mobile client. The main reason for this choice is that HTTP is the de-facto transport protocol for Web services ranging from file sharing to media streaming, and the studies analyzing the Internet traffic~\cite{Mirkovic2015} show that HTTP accounts for 75\% of global mobile Internet traffic.

\begin{figure}[t]
	\captionsetup{justification=centering}
	\begin{center}
		\includegraphics[width=90mm, height=80mm]{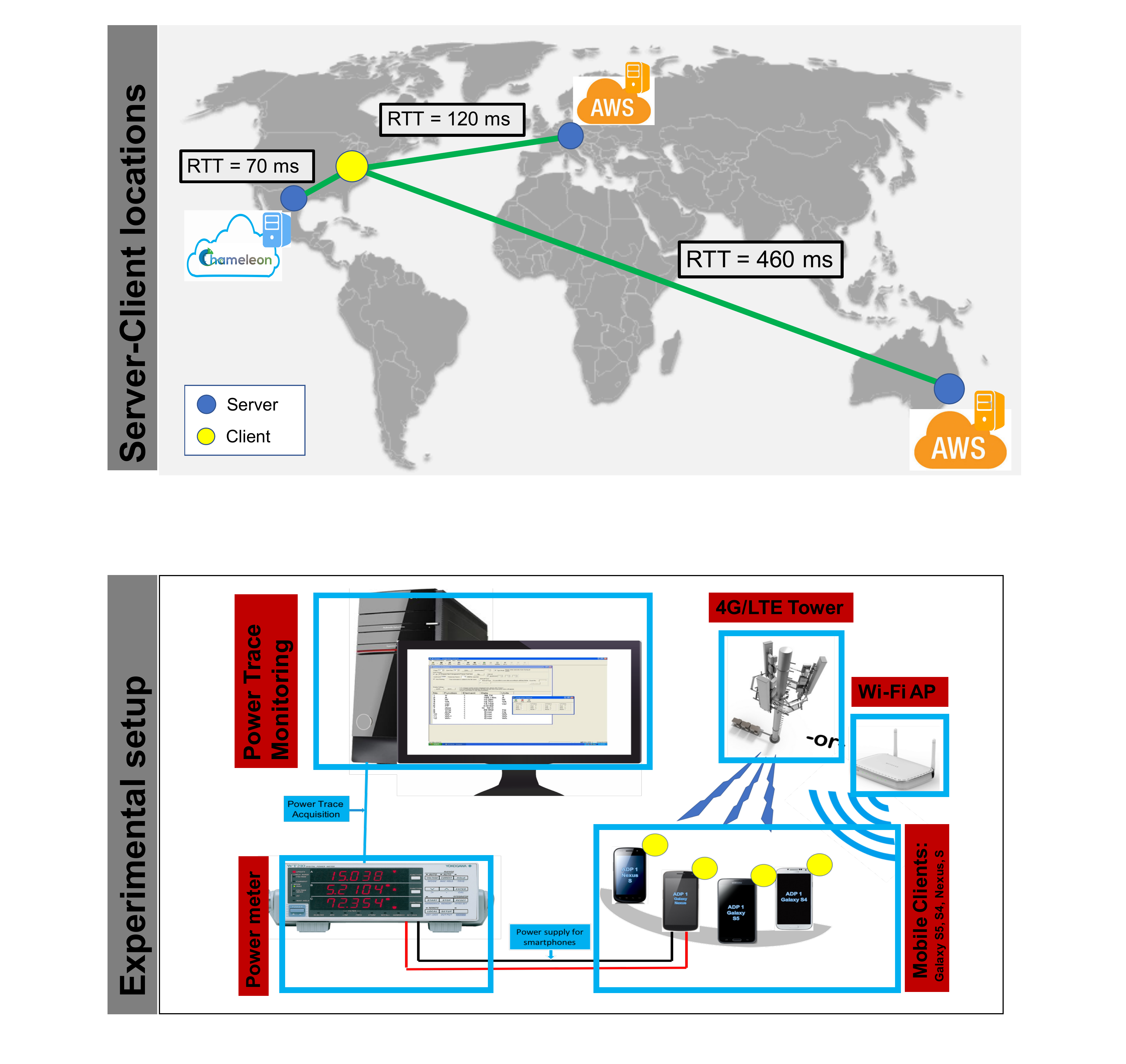}
		\vspace{-5mm}
		\caption{Network map of the experimental testbed and the setup of the power measurement system.}
		\label{fig:exSetup}
	\end{center}
\end{figure}

We analyzed the data transfer throughput of HTTP data transfers and the power consumption during which we run tests with different level of concurrency (cc), parallelism (p), and combined concurrency \& parallelism parameters. We also measured the instantaneous power consumption and total energy consumption of each individual request among different web servers and clients.
The experiments were conducted on Amazon Elastic Compute Cloud (AWS EC2) instances, Chameleon Cloud~\cite{chameleon}, and Data Intensive Distributed Computing Laboratory (DIDCLAB). The network map of the experimental testbed and the setup of the power measurement system are illustrated in Figure~\ref{fig:exSetup}.

We used varying size of files (HTML, image, and video) to analyze the effect of each tuning parameter on transfer throughput and energy consumption. The characteristics of these files are presented in Table~\ref{tab:dataset}. 
In order to increase the robustness of the obtained throughput and energy consumption values for each experimental setting, we run each test within the range of five to ten times, and the average values of throughput and energy consumption were used. As a result of iteration of each individual experiment among four different mobile clients and three different web servers with different bandwidth (BW) and round-trip-time (RTT), we transferred varying size of nearly 3.8 Million individual files.

\begin{table}[t]
	\small
	\centering
	\caption{Characteristics of the dataset used in the experiments.} \label{tab:dataset}
	\begin{tabular}{| r@{\hskip 0.5cm}| r@{\hskip 0.5cm} | r |}
		\hline
		\rule{0pt}{3ex}
		{\bf Dataset Name} &  {\bf Ave. File Size}  & {\bf Min - Max } \\
		\hline
		\hline
		\rule{0pt}{2.3ex}
		HTML	&  112 KB 	& 56 KB - 155 KB 	\\
		\hline
		\rule{0pt}{2.3ex}
		IMAGE	&  2.7 MB 	& 2 MB - 3.2 MB 	\\
		\hline
		\rule{0pt}{2.3ex}
		VIDEO	&  152 MB 	& 140 MB - 167 MB 	\\
		\hline
	\end{tabular}
	\vspace{-3mm}
\end{table}

\begin{figure}[t]
	\begin{centering}
		\captionsetup{justification=centering}
		\begin{tabular}{cc}
\hspace{-0.45cm}			\includegraphics[keepaspectratio=true,angle=0,width=45mm]{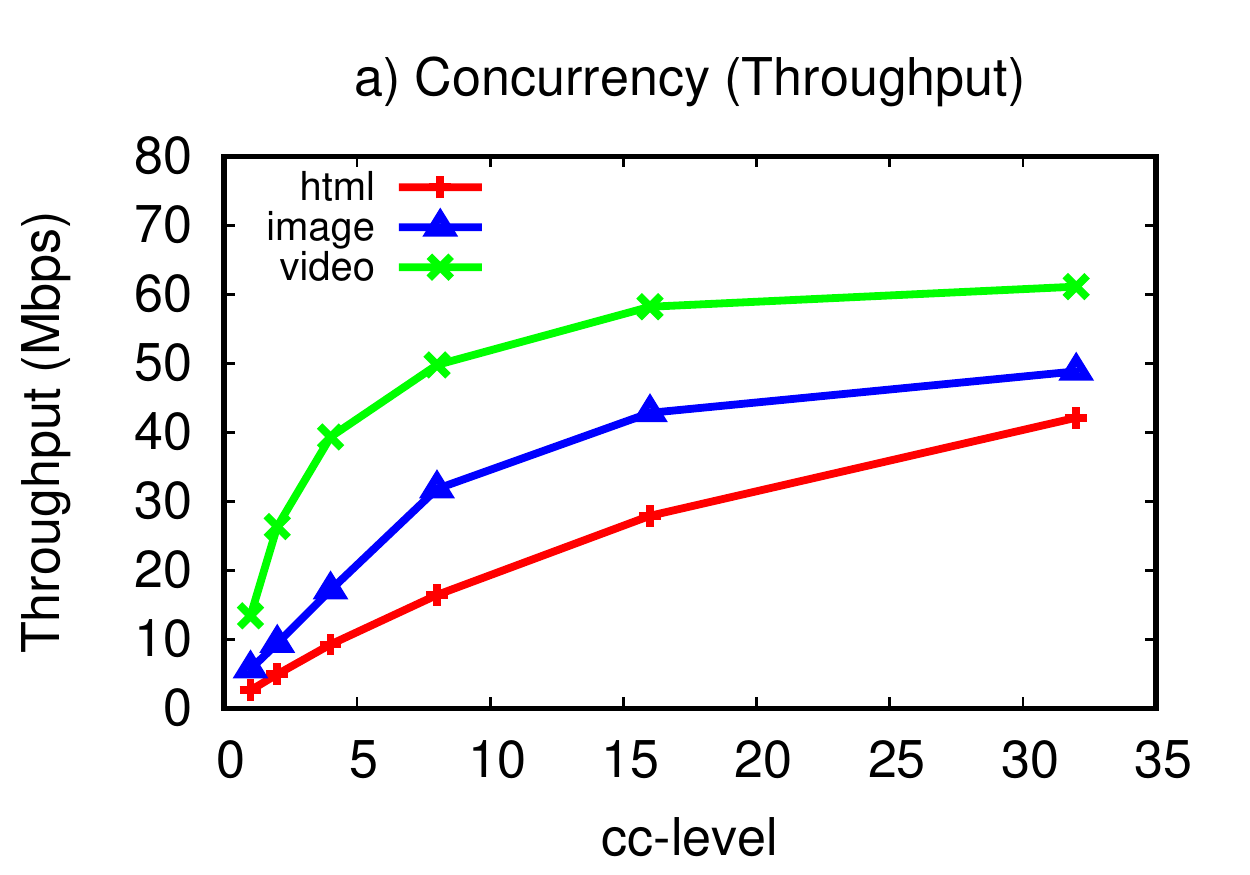}
			\includegraphics[keepaspectratio=true,angle=0,width=45mm]{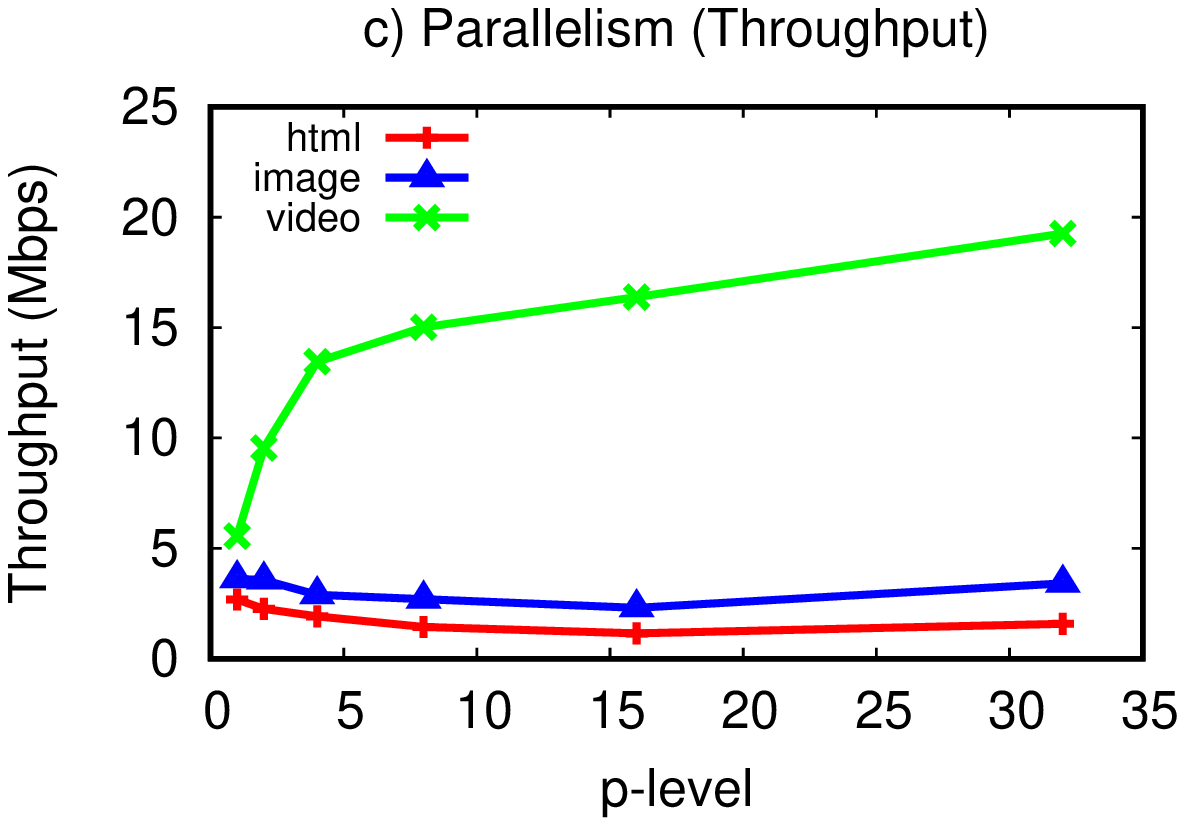}\\
\hspace{-0.45cm}			\includegraphics[keepaspectratio=true,angle=0,width=45mm]{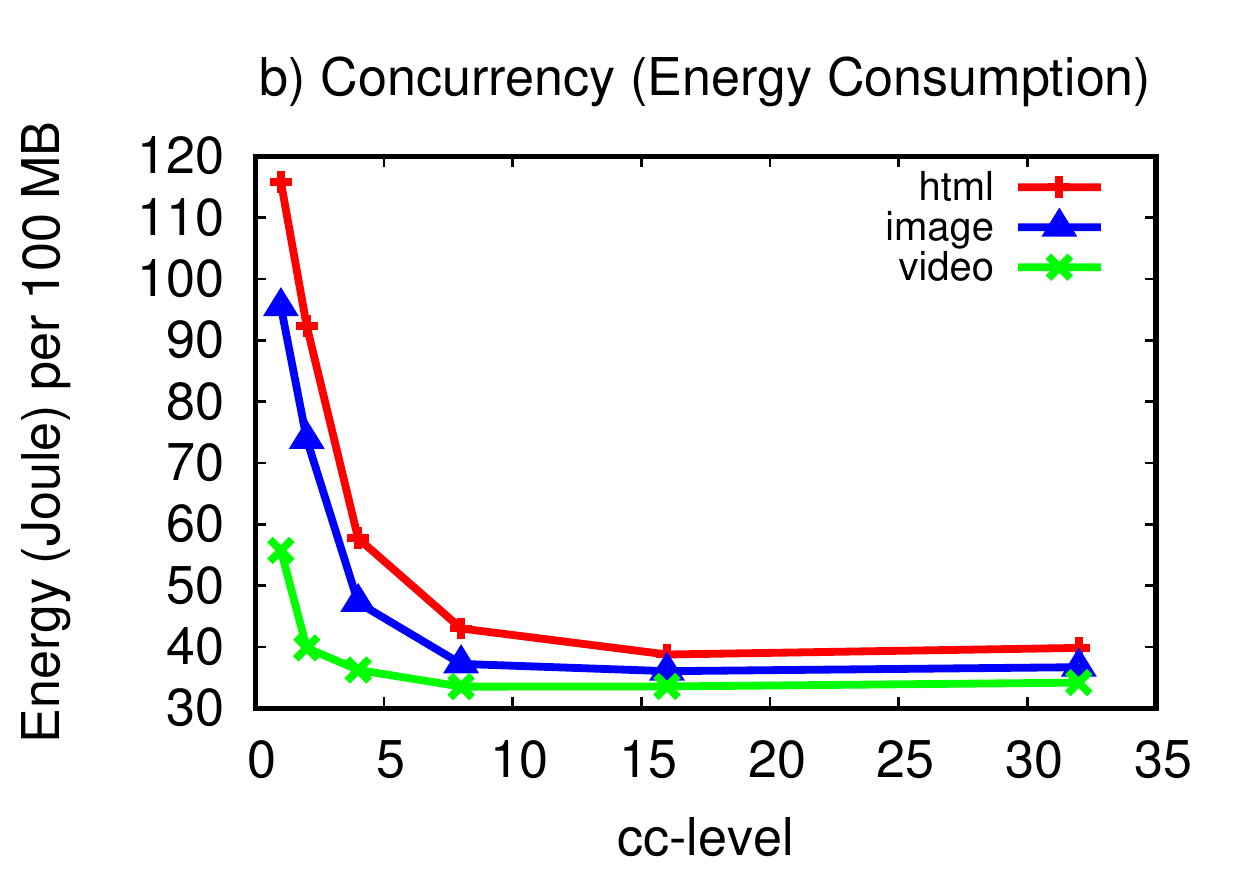}
			\includegraphics[keepaspectratio=true,angle=0,width=45mm]{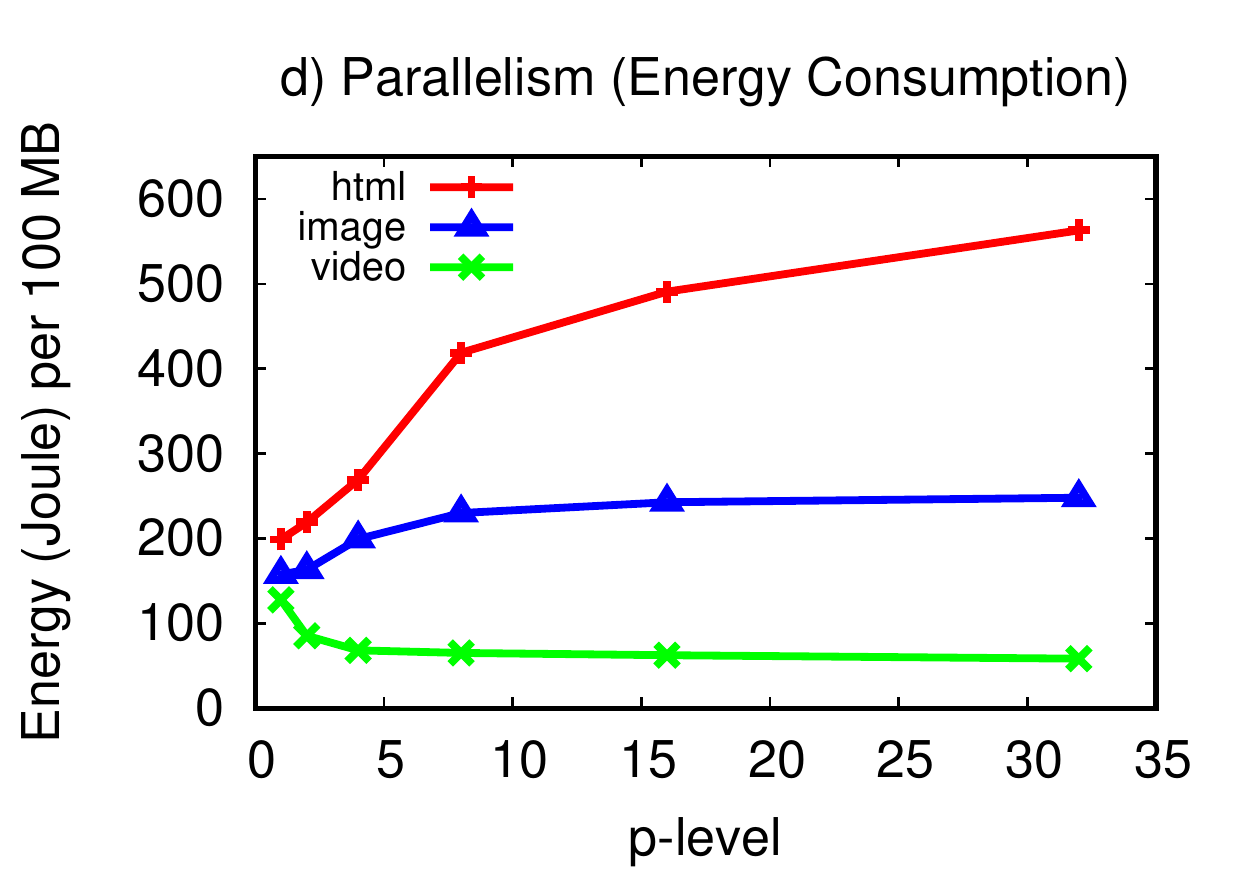}
		\end{tabular}
		\caption{Throughput vs Energy Consumption trade-offs of individual protocol parameters for WiFi data transfers between AWS EC2 Sydney and DIDCLAB Galaxy S5.} \label{fig:parameter}
	\end{centering}
\end{figure}

\begin{figure*}[t]
	\captionsetup{justification=centering}
	\begin{centering}
		\begin{tabular}{cc}
			\includegraphics[keepaspectratio=true,angle=0,width=52mm]{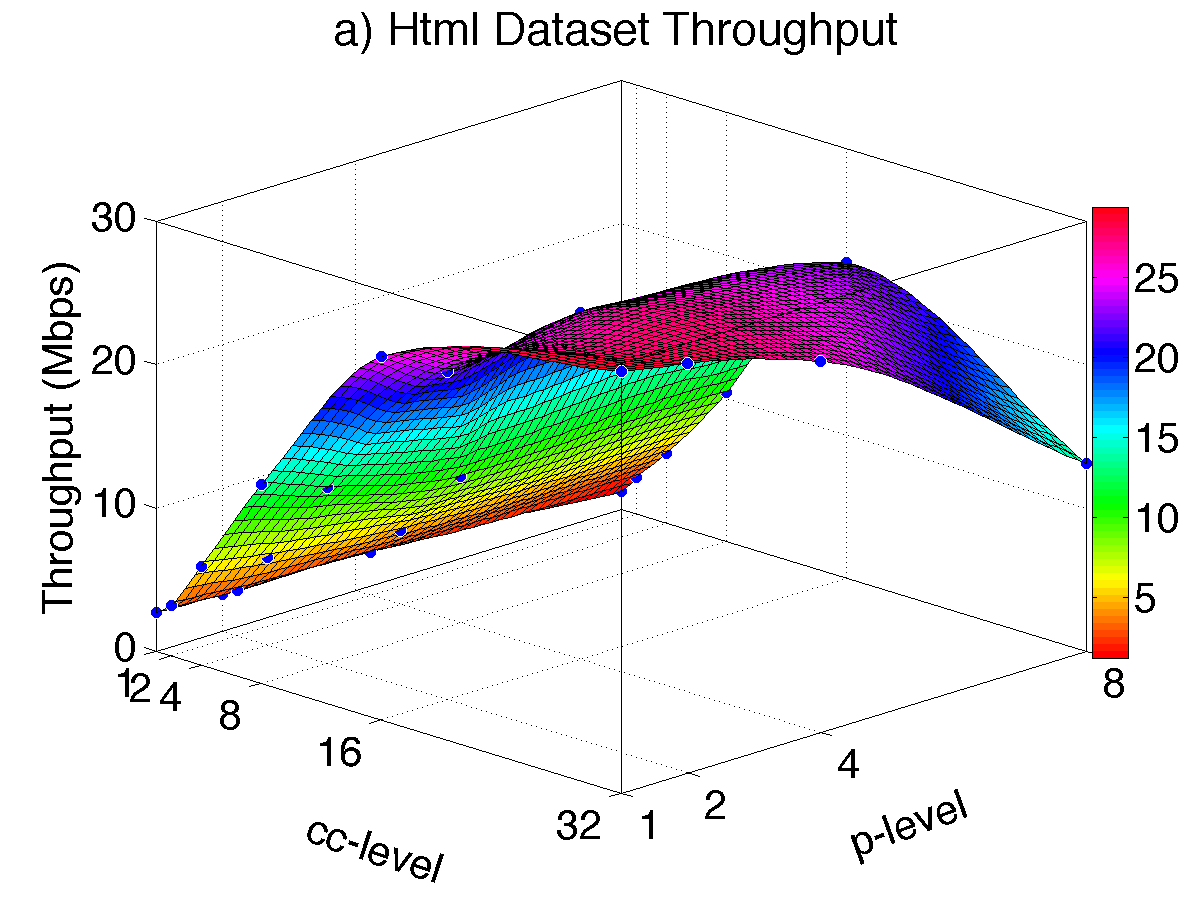}
			
			\includegraphics[keepaspectratio=true,angle=0,width=52mm]{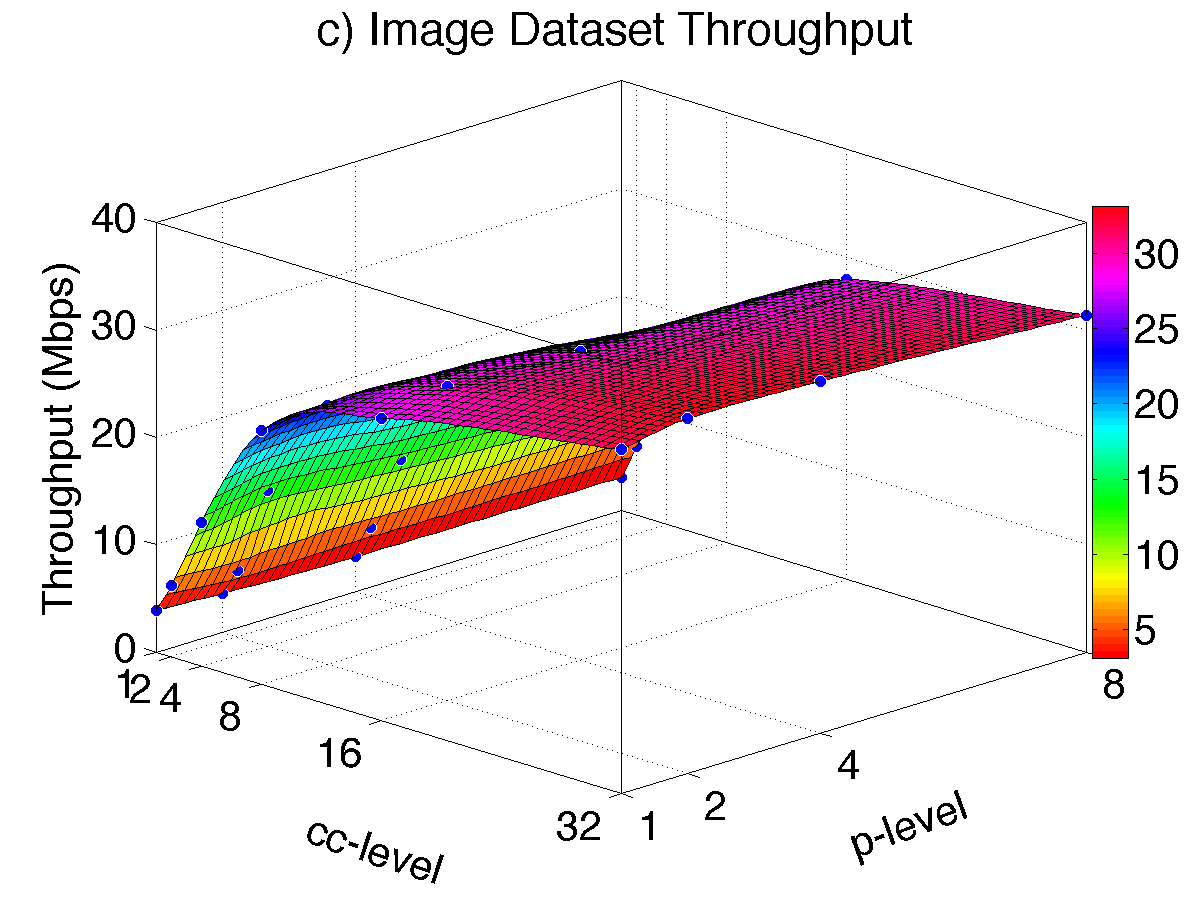}
			
			\includegraphics[keepaspectratio=true,angle=0,width=52mm]{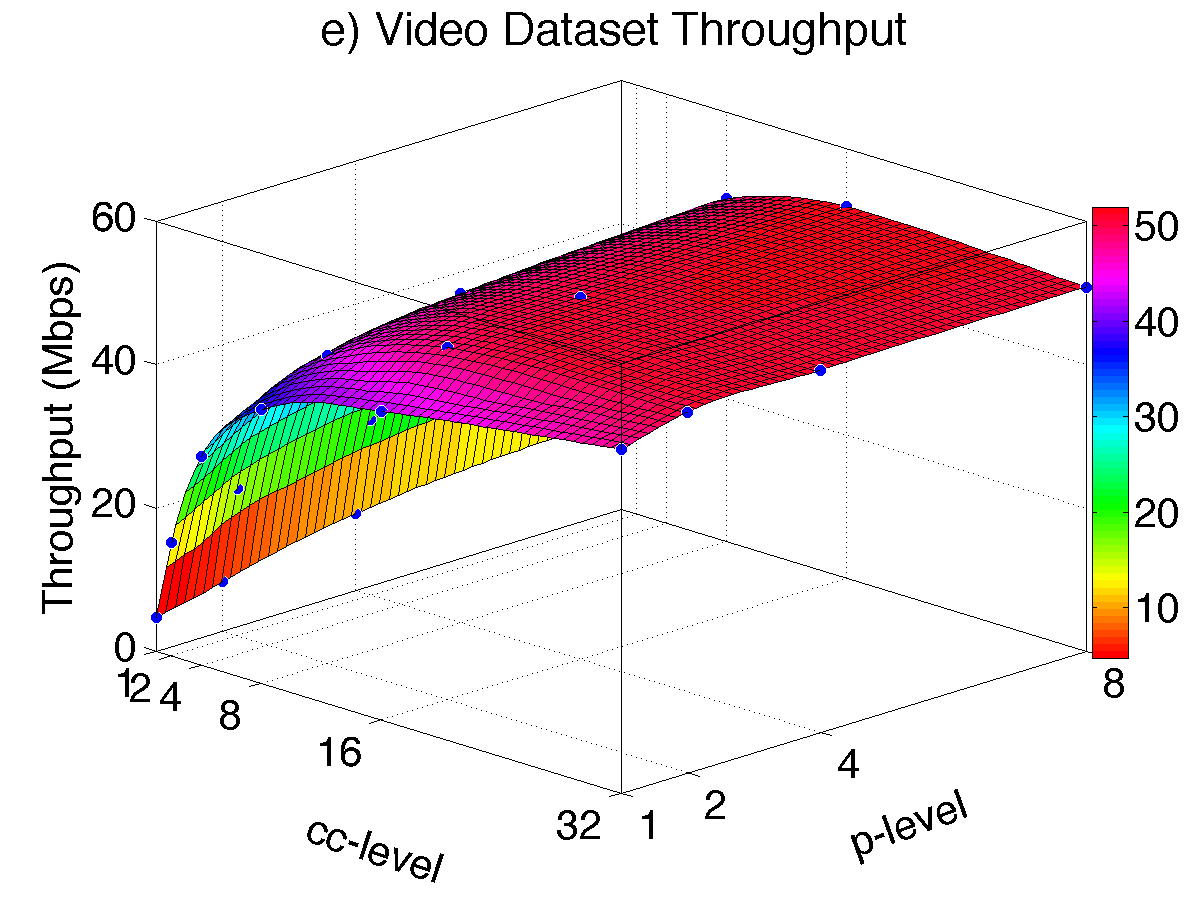}\\
			
			\includegraphics[keepaspectratio=true,angle=0,width=52mm]{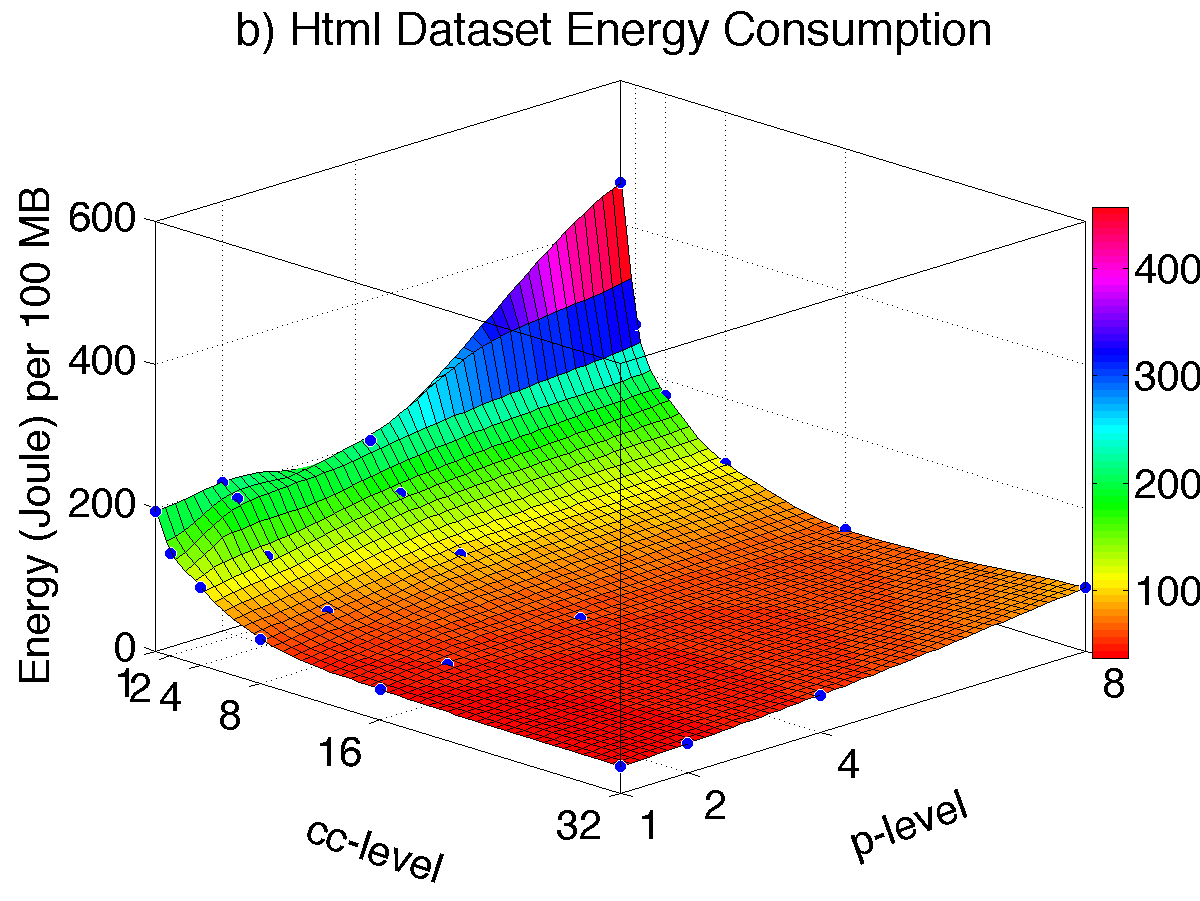}
			\includegraphics[keepaspectratio=true,angle=0,width=52mm]{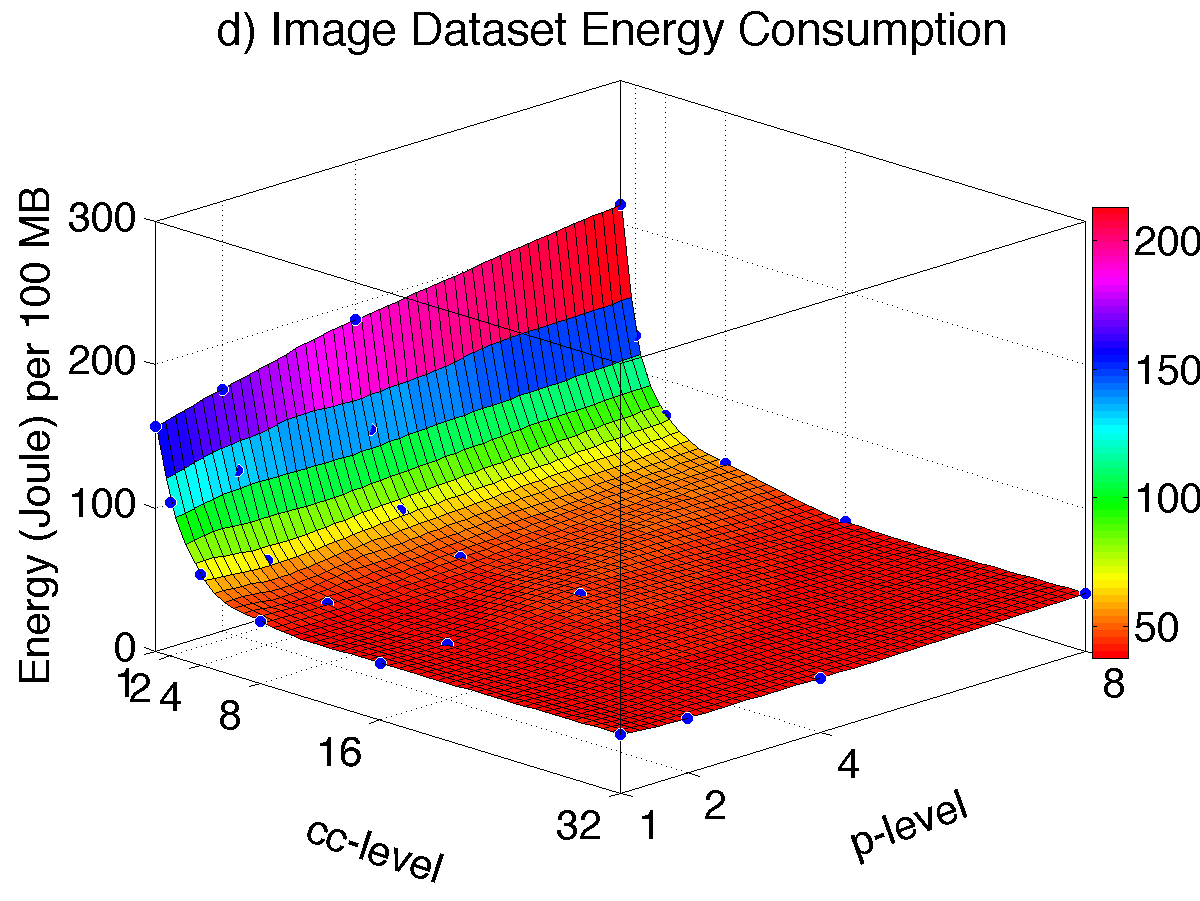}
			\includegraphics[keepaspectratio=true,angle=0,width=52mm]{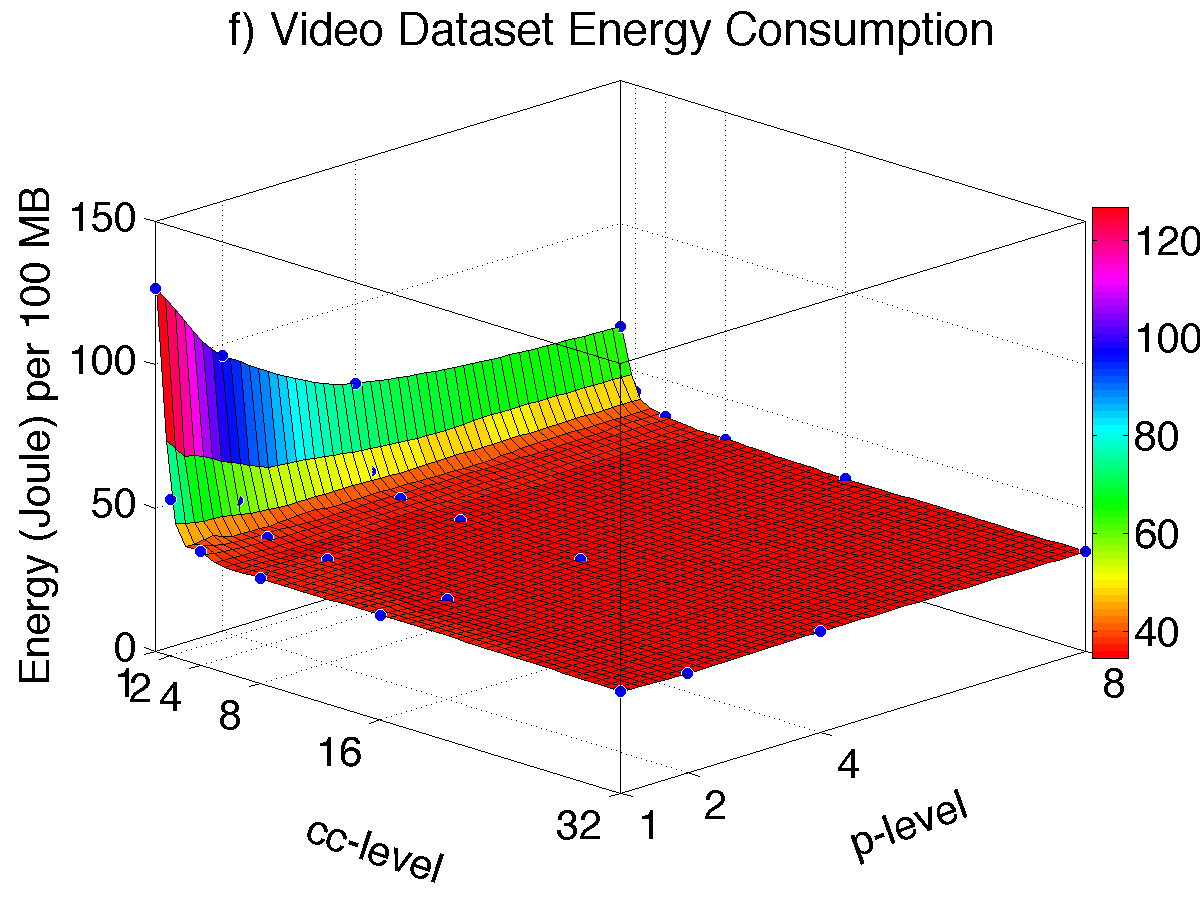}
			
		\end{tabular}
		\caption{Throughput vs Energy Consumption trade-offs of combined protocol parameters for WiFi data transfers from AWS EC2 Sydney to DIDCLAB Galaxy S5.} \label{fig:matlabS5}
	\end{centering}
\end{figure*}

Initially, we tested the individual parameter effects of concurrency and parallelism on the achieved throughput and energy consumption for the data transfers between the web server at AWS EC2 Sydney and the client Samsung Galaxy S5 at DIDCLAB in Buffalo. Overall, concurrency parameter showed a better performance than parallelism on HTML, image, and video file transfers. When we increased level of concurrency from 1 to 32, it boosted end-to-end throughput for all all file types and reduced energy consumption on the mobile client as seen in Figure~\ref{fig:parameter}(a). On the other hand, when it comes to the parallelism parameter, the performance of each file type showed different characteristics. Increased level of parallelism improved the end-to-end throughput of the video file transfers and decreased the energy consumption up to a specific level as shown at Figure~\ref{fig:parameter}(b). These results show that parallelism is especially effective during large file transfers.

Having throughput and energy consumption results of individual parameters on data transfers, we designed a simple download manager that uses the combination of concurrency and parallelism parameters in order to increase throughput while saving energy. Figure~\ref{fig:matlabS5} shows throughput versus energy consumption (per 100 MB) trade-offs of our application level parameters on the same datasets from AWS EC2 Sydney to Galaxy S5 at DIDCLAB in Buffalo. When concurrency level increased from 1 to 32 as well as parallelism from 1 to 8, throughput slightly improved for the html and image datasets compared to individual parameter results in terms of parallelism as seen in Figure~\ref{fig:matlabS5}(a)-(d) and energy consumption per 100 MB increased when the level of parallelism increased on fixed concurrency level. On the other hand, concurrency still managed to show its positive effect on throughput at each parallelism level. Parallelism became more effective for larger file transfer as expected (Figure~\ref{fig:matlabS5}(e)-(h)). 

Overall, using the combined parameters, we managed to increase the highest energy saving result of individual parameters up to 81\%. We run the same experiments with other smartphones as well: while Galaxy S4 presented similar but less throughput and higher energy efficiency, Galaxy Nexus N3 and Nexus S showed moderate performance compared to Galaxy S5.  Due to the space limitations of the paper, we had to limit the number of graphs we can present.

\section{Proposed Algorithms}
\label{sec:algorithms}

The guaranteed reliability and quality of the provided services become more important according to the end-users' needs. Hence, we have developed three different data transfer optimization algorithms for mobile users based on our initial analyzes. These include: (1) Lowest-possible Energy (LowE); (2) Highest-achievable Throughput (HAT); and (3) Energy-aware High Throughput (EHT) algorithms. 

\begin{algorithm}[b!]  
\scriptsize
\centering
\caption{--- Lowest-possible Energy Algorithm} \label{alg:LowE}
\begin{flushleft}
\textbf{\normalsize Input\quad : }{RTT, BW, maxCC, allFileList}\\
\textbf{\normalsize Output : }{Optimal\ Transfer\ Parameters\ for LowE}
\vspace{-0.3cm}
\end{flushleft}
\begin{spacing}{1}
\begin{algorithmic}[1]
  \Statex
    \Function{LowE Data Transfer}{maxCC, BW, RTT}
    \State $maxCC \leftarrow availChannel$
	\State$BDP \leftarrow \ BW * RTT$
    \State $allFileList \leftarrow fetchFileListFromWebServer()$
	\State $ C \leftarrow cluster(allFileList)$
    \For{$\textbf{each} \ c_{i} \ \textbf{in} \ C$}
        \State$avgFileSize \leftarrow computeAverage(c_{i})$
        \State$cc_{opt} \leftarrow \ optParams(BDP, avgFileSize, availChannel)$ \label{line:concurrency}
         \State$p_{opt} \leftarrow \ optParams(BDP, bufSize, availChannel)$ \label{line:parallelism} 
     \While{$(availChannel \neq 0)$}
		\State$availChannel \leftarrow maxCC - assignedChannels$
     \EndWhile
    \EndFor
\State $startTransferForAll(c_{i})$ \label{LowE:transfer}
	\EndFunction
\end{algorithmic}
\end{spacing}
\end{algorithm}

The {\bf Lowest-possible Energy (LowE)} algorithm aims to achieve the minimum energy consumption during dowdload/upload of data transfer by tuning application-layer network parameters mentioned earlier. The main goal of LowE algorithm is minimizing energy consumption with no concern on performance, which brings flexibility on transfer completion time.
Thus, our application-layer solution, which do not require any lower-layer protocol change, can be used data syncing/backing up in cloud computing, background transfers, updates,etc.
We initially divide dataset into different number of chunks based on the characteristics of the dataset and network used, and then treat each chunk separately. After dataset is divided into chunks, we calculate optimal concurrency level ($cc_{opt}$) based on the bandwidth-delay-product ($BDP$), average file size for each chunk, and number of available channels. Optimal parallelism level ($p_{opt}$) is based on TCP buffer size ($bufSize$), BDP, and number of available channels (line 8 and 9). $BDP$ is calculated as a product of the bandwidth of the network link ($BW$) and round-trip-time ($RTT$) (line 3). The details of the LowE algorithm are shown in Algorithm~\ref{alg:LowE}. 

\begin{algorithm}[hbt]  
\scriptsize
\centering
\caption{--- Highest-achievable Throughput Algorithm} \label{alg:HAT}
\begin{flushleft}
\textbf{\normalsize Input\quad : }{RTT, BW, maxCC, allFileList}\\
\textbf{\normalsize Output : }{Optimal\ Transfer\ Parameters\ for\ HaT }
\vspace{-0.3cm}
\end{flushleft}
\begin{spacing}{1}    
\begin{algorithmic}[1]  
  \Statex
    \Function{HAT Data Transfer}{maxCC, BW, RTT}
    \State $maxCC \leftarrow availChannel$
	\State $BDP \leftarrow BW * RTT$
	\State $allFileList \leftarrow fetchFileListFromWebServer()$
	\State $ C \leftarrow cluster(allFileList)$ 
    \State $findOptimalParameters()$
	\While{$(availChannel \neq 0)$}
    	\State $i$ = $(m * 3)$ $+$ $r$ \Comment{m (m $\geq$ 0) is quotient of i}
    	\For { $i \leftarrow 1$ to $ maxCC $}
        \State $i \leftarrow r \pmod 3$	
			\If{$(r$==$1)$}
				\State $cc \ $+=$ \ $m+1$ \ for \ (c_{video})$ \Comment{$cc \leftarrow \textnormal{concurrency}$}
                \State $cc \ $+=$ \ $m$ \ \quad for \ (c_{image},c_{html})$
			\ElsIf{$(r$==$2)$}
				\State $cc \ $+=$ \ $m+1$ \ for \ (c_{video},c_{image})$
                \State $ cc \ $+=$ \ $m$ \ \quad for \ (c_{html})$
			\Else
				\State $cc \ $+=$ \ $m$ \ \quad for \ (c_{video},c_{image},c_{html})$
			\EndIf
		\EndFor
    \EndWhile    
	\State $startTransferForAll(c_{i})$ \label{HAT:transfer}
	\EndFunction
	\end{algorithmic} 
\end{spacing}
\end{algorithm}

The {\bf Highest-achievable Throughput (HAT)} algorithm, on the other hand, aims to maximize the throughput of data transfer without energy concerns. Since HAT focuses on completing the data transfer at closest time possible, it can be used for time sensitive applications such as real-time audio/video streaming and interactive games. Similar to LowE, HAT partitions files into three chunks (small, medium and large) according to their size. Then it calls $findOptimalPArameters()$ function to calculate the best possible optimal parameter values for application-layer network parameters (line 6 in Algorithm~\ref{alg:HAT}), and finally HAT assigns channels to chunks using round-robin algorithm in the order of large $>$ medium $>$ small as seen through line 7 to line 21 in Algorithm~\ref{alg:HAT}. The algorithm prioritize by assigning higher concurrency values for large, medium and small chunks, respectively.

\begin{algorithm}[b!]
\scriptsize
\centering
\caption{--- Energy-aware High Throughput Algorithm} \label{alg:EHT}
\begin{flushleft}
\textbf{\normalsize Input\quad : }{RTT, BW, maxCC, allFileList, efficientCC} \\
\textbf{\normalsize Output : }{Optimal\ Transfer\ Parameters\ for\ EHT}
\vspace{-0.3cm}
\end{flushleft}
\begin{spacing}{1} 
\begin{algorithmic}[1] 
  \Statex
    \Function{EHT Data Transfer}{maxCC, BW, RTT}
	\State $BDP \leftarrow BW * RTT$
	\State $allFileList \leftarrow fetchFileListFromWebServer()$
	\State $ C \leftarrow cluster(allFileList)$
    \State $findOptimalParameters()$
    \For{$\textbf{each} \ cluster\ (c_{i}) \ \textbf{in} \ C$} \Comment{$c_{i} \in C\ for\ $i=0...N$ (N \geq 0)$}
		\State $cluster.weight \leftarrow cluster.totalSize * cluster.fileCount$
        \State $TotalWeight\ $+=$\ cluster.weight$
    \EndFor
	\For{$\textbf{each} \ cluster\ (c_{i}) \ \textbf{in} \ C$} \label{ea:initialAllocation}
    	\State $cluster.weight \leftarrow {cluster.weight} \ /\ {TotalWeight}$
		\State $cluster.channelAllocation \leftarrow \left\lfloor maxCC * cluster.weight \right\rfloor $
	\EndFor
	\State $energyEfficiency[]$;
    \For{$i=1; \ i \leq maxCC; \ i$+=$3$} \Comment{$actChannel \leftarrow i$}
		\State $startTransfer(actChannel)$		
		\State $energyConsumption=calculatePowerMetrics()$
		\State $throughput=calculateThroughput()$
		\State $energyEfficiency[i]= \frac{throughput}{energyConsumption}$
	\EndFor
	\State $efficentCC=max(energyEfficiency[ ])$
	\State $transfer(efficentCC)$ \label{EHT:transfer}
\EndFunction
\end{algorithmic}
\end{spacing}
\end{algorithm}

In the {\bf Energy-aware High Throughput (EHT)} algorithm, our goal is to balance throughput vs energy consumption rate by finding best optimal parameter configuration to achieve the maximum throughput on the given networks as well as minimizing energy consumption at the same time. ETH algorithm differs from LowE and HAT algorithms in terms of taking into consideration weights when it comes to assigning channels (instead of using simple heuristic approach). Additionally, it examined many values to reach maximum $throughput/energy$ ratio while others use pre-calculated concurrency levels.
Similar to HAT, EHT first partitions files into chunks, then calculates optimal pipelining and parallelism levels. The focus of the algorithm is not only to get maximum throughput for each chunk or to minimize energy consumption to the lowest possible level. Instead it is designed to increase performance for each chunk as well as to decrease energy consumption at the same time within the available channel range by using concurrency parameter. For channel allocation, our EHT algorithm takes into account weights when it comes to assigning threads to each chunk. The minimum value of concurrent level of available channels is one for all clusters and the maximum value is assigned based on resource capacities of mobile users and fairness concerns. 

\section{Evaluation of the Algorithms}
\label{sec:evaluation}

\begin{figure*}[t]
	\captionsetup{justification=centering}
    \begin{centering}
		\hspace*{-0.2cm}\begin{tabular}{ccc}

\includegraphics[keepaspectratio=true,angle=0,width=58mm]{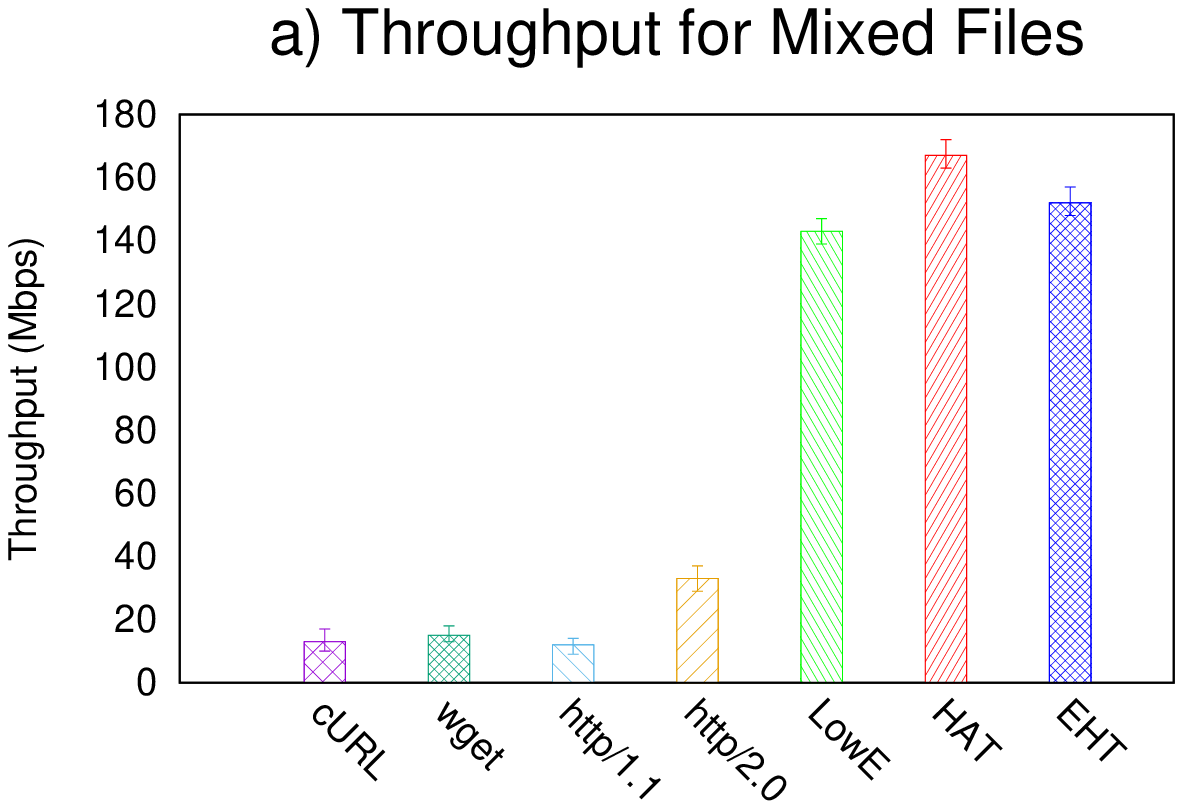}
			\includegraphics[keepaspectratio=true,angle=0,width=58mm]{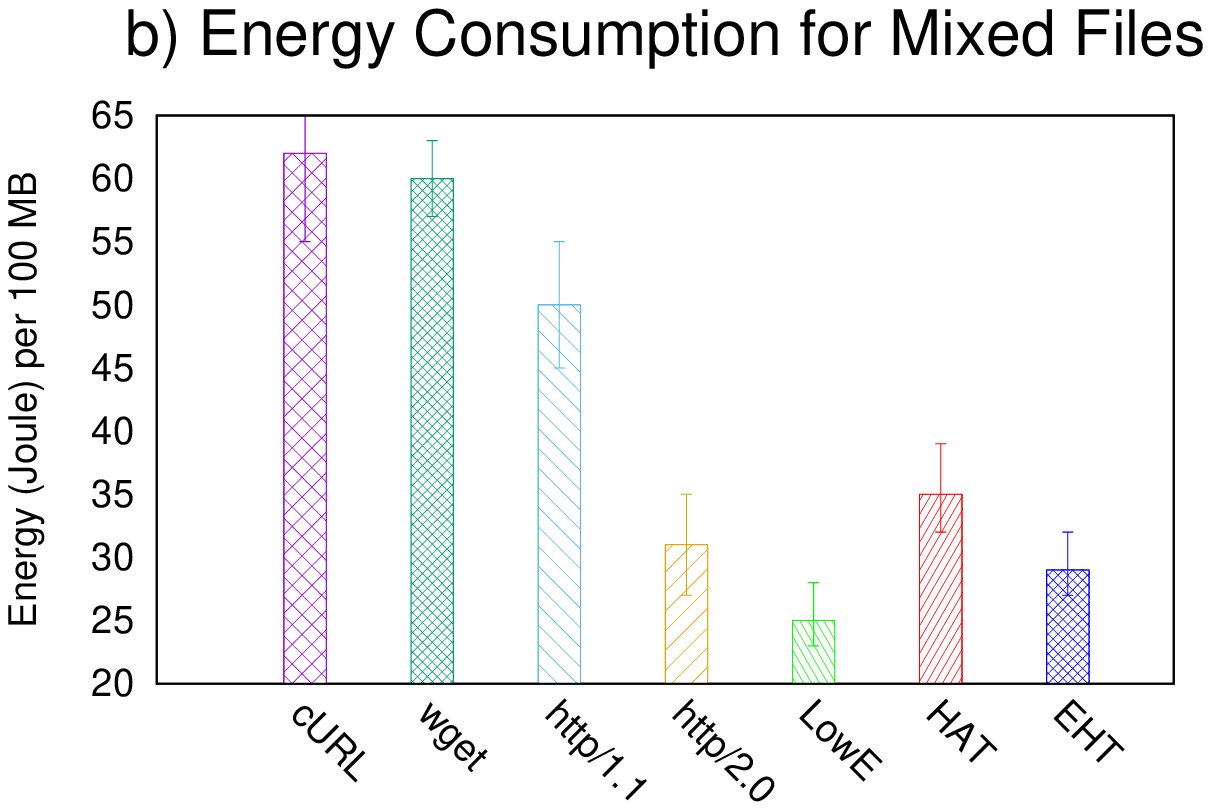}
			\includegraphics[keepaspectratio=true,angle=0,width=58mm]{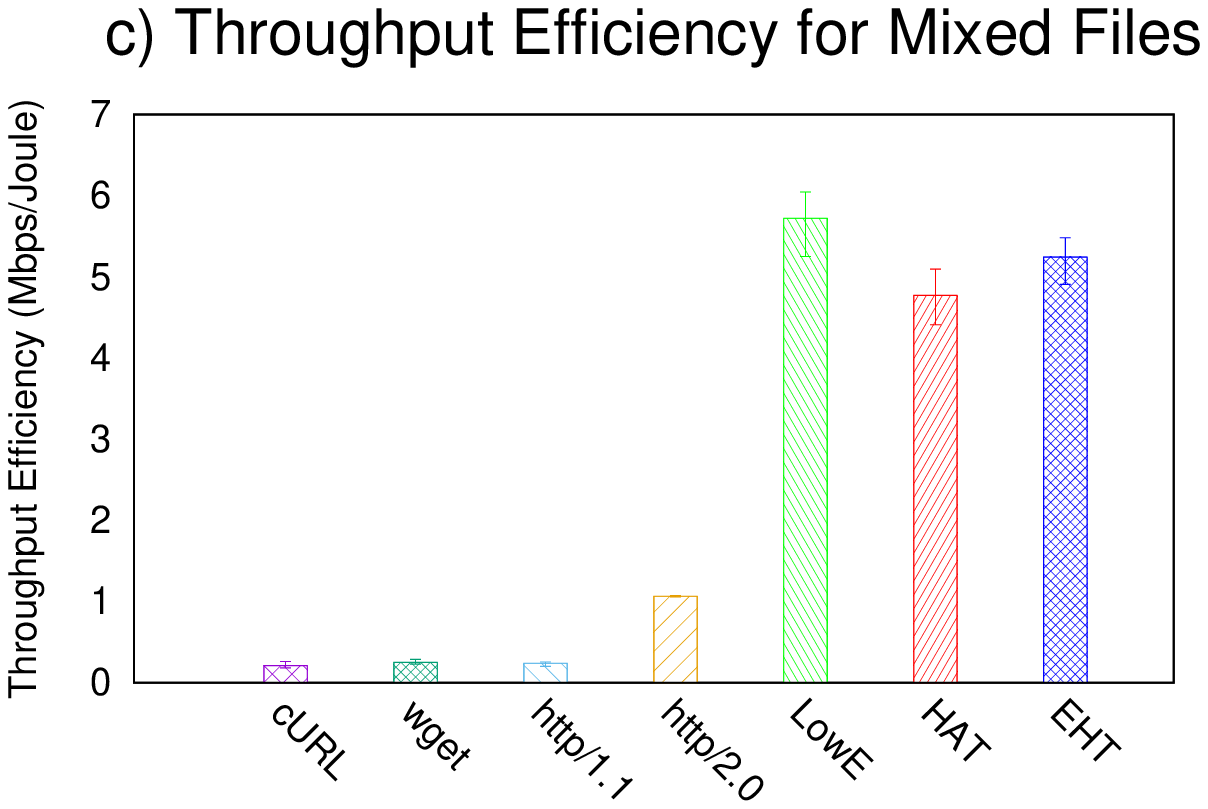}
		\end{tabular}
		\caption{Maximum achieved throughput, corresponding energy consumption, and throughput efficiency of the proposed algorithms compared with state-of-the-art.} 
        \label{fig:algResults}
	\end{centering}
\end{figure*}

We compared the performance and power consumption of our three algorithms (LowE, HAT and EHT) with energy-agnostic wget~\cite{wget} and curl~\cite{curl} clients as well as two different versions of the de-facto application layer transfer protocol of HTTP, which are HTTP/1.1 and HTTP/2. The newly introduced HTTP/2 is superior to HTTP/1.1 in terms of being a binary protocol that supports multiplexing, header compressions and letting the server to {\em push} responses. We used a combination of HTML, image and video datasets to compare our LowE, HAT and EHT algorithms with other methods/models in order to get a fine-grained analysis of performance and energy consumption. 

Figure~\ref{fig:algResults} shows both throughput performance and energy consumption of different models. As seen in Figure~\ref{fig:algResults}(a) our all three algorithms outperform other tested solutions in terms of throughput gain. The performance improvement is approximately $\times$ over the closest competitor, HTTP/2. Even though HTTP/2 uses multiplexing to allow multiple requests and response over a single connection for head-of-line blocking issue, it still obtains very poor results over the network links with high latency. On the other hand, while HTTP/1.1 allows multiple connections, it also lacks of dynamically tuning number of connections. 

As shown in Figure~\ref{fig:algResults}(b) energy-agnostic applications wget and cURL consume 2.7$\times$ and 2$\times$ times more energy compared with our LowE algorithm and HTTP/2, respectively. We also observed that when only smaller files are transfered with wget and cURL, the energy consumption rate even increases drastically. 
Figure~\ref{fig:algResults}(c) shows the throughput efficiency of different models. We use Equation \ref{eq:thEff} to calculate throughput efficiency of our algorithms and other methods.

\vspace{-0.5cm}
\begin{equation}
Throughput\_efficiency = \frac{throughput}{energy\_consumed}
\label{eq:thEff}
\end{equation}

As seen in Figure~\ref{fig:algResults}(c) newly introduced HTTP/2 nearly 4$\times$ times throughput efficient comparing with widely used HTTP/1.1. On the other hand, the EHT algorithm outperforms its closest competitor HTTP/2 over 5.2$\times$ times.

\section{Conclusion}
\label{sec:conclusion}

In this paper, we performed extensive analysis on the effects of application-layer data transfer protocol parameters (such as the number of parallel data streams per file, and the level of concurrent file transfers to fill the mobile network pipes) on mobile data transfer throughput and energy consumption for WiFi and 4G LTE connections. 

Based on our analysis results, we proposed three novel application-layer algorithms (i.e., LowE, HAT, and EHT) for wireless networks to increase energy saving rates of mobile users without sacrificing throughput performance. Our LowE, HAT, and EHT algorithms show that significant energy savings can be achieved with application-layer solutions at the mobile systems during data transfer with no  performance penalty. We also show that, in many cases, performance increase and energy savings can be achieved simultaneously.

In the experiments, we show that by only tuning application-layer parameters (i.e., concurrency and  parallelism) during data transfers, an energy saving up to 2.7$\times$ achieved by our LowE algorithm. At the same time, high throughput gain of the end-to-end data transfer obtained compared with standard applications like cURL and wget. However, our HAT algorithm achieves the highest overall throughput gain. Additionally, our energy-efficient EHT algorithm outperforms its closest competitor HTTP/2 by up to 5.2$\times$ times.

\bibliographystyle{IEEEtran}
\bibliography{main}

\end{document}